\documentstyle[12pt]{article}
\def\be{\begin{equation}}
\def\ee{\end{equation}}
\def\ba{\begin{array}{c}}
\def\ea{\end{array}}
\def\p{\partial}
\def\ben{\[}
\def\een{\]}
\begin{document}

\titlepage

  \begin{center}

\section* {\Large \bf
Multiparametric oscillator Hamiltonians with exact bound states in
infinite-dimensional space
 }

\end{center}

\vspace{5mm}

   \begin{center}

Miloslav Znojil

\vspace{3mm}

{\small \it \'{U}stav jadern\'e fyziky AV \v{C}R, 250 68 \v{R}e\v{z},
Czech Republic\\

e-mail: znojil@ujf.cas.cz}

\vspace{5mm}

\end{center}

\subsection*{Abstract}

Bound states in quantum mechanics must almost always be
constructed numerically. One of the best known exceptions concerns
the central $D-$dimensional (often called ``anharmonic")
Hamiltonian $H = p^2 + a\,|\vec{r}|^2 + b\,|\vec{r}|^4 + \ldots +
z\,|\vec{r}|^{4q+2}$ (where $z=1$) with a complete and elementary
solvability at $q=0$ (central harmonic oscillator, no free
parameters) and with an incomplete, $N-$level elementary analytic
solvability at $q=1$ (so called ``quasi-exact" sextic oscillator
containing one free parameter). In the limit $D \to \infty$,
numerical experiments revealed recently a highly unexpected
existence of a new broad class of the $q-$parametric quasi-exact
solutions at the next integers $q=2, 3, 4$ and $q=5$. Here we show
how a systematic construction of the latter, ``privileged" $D \gg
1$ exact bound states may be extended to much higher $q$s (meaning
an enhanced flexibility of the shape of the force) at a cost of
narrowing the set of wavefunctions (with $N$ restricted to the
first few non-negative integers). At $q=4K+3$ we conjecture a
closed formula for the $N=3$ solution at all $K$.

\vspace{5mm}



MSC 2000: 81Q05 13P05 14M12

\newpage
\section{Introduction}

In quantum mechanics, bound states of a particle confined in a
central potential well $V(|\vec{r}|)$ in $D$ dimensions are
constructed as normalizable solutions of the ordinary differential
Schr\"{o}dinger equation
 \be
 \left[-\,\frac{d^2}{dr^2} + \frac{\ell(\ell+1)}{r^2} +
 V^{}(r)
 \right]\, \psi(r) =
 E\, \psi(r)\,.
 \label{rad}\
  \ee
Usually, this equation (or, more precisely, their infinite set
numbered by integer argument $ L = 0, 1, \ldots$ of the ``angular
momentum" $ \ell = \ell(L)= L + ({D-3})/{2}$) must be solved by
purely numerical means.

A notable exception concerns all the models where the spatial
dimension $D$ proves ``sufficiently" large, $D \gg 1$. In a way
outlined, say, in reviews \cite{Bjerrum}, a semi-analytic,
perturbative construction of the solutions may be then based on
the so called large$-D$ expansion technique. In essence, this
approach combines our knowledge of the asymptotic growth of $V(r)
\approx r^\alpha$ at $ r \gg 1$ with the presence of the strong
repulsive core $\ell(\ell+1)/r^2$ near the origin. This implies
the existence of a (presumably, pronounced) absolute minimum of
the combined forces at a point $r = R(\alpha)$ where we have
 \ben
 \p_y\left .\left [
\frac{\ell(\ell+1)}{(R+y)^2}+ (R+y)^\alpha
 \right ] \right |_{y=0}=0, \ \ \ \ \ \ \Longrightarrow
  \ \ \ \ \ \ \
 R=R(\alpha) =
 \left [
 \frac{2\ell(\ell+1))}{\alpha}
 \right ]^{1/(\alpha+2)} \gg 1\,
 \een
whenever we neglect all the less relevant corrections. The main
merit of this approach lies in its simplicity and universality
but, unfortunately, its expected rate of convergence (measured by
the smallness of $1/R$) will, obviously, increase quite slowly
with the growth of $D$ and decrease quite quickly with an increase
of the dominant power of the potential $\alpha$. As long as we are
going to pay attention to the class of potentials
 \be
V_{(q,k)}(r) =\frac{1}{r^2} \,\left [
g_{-2}+g_{-1}\,r^{2/(k+1)}+g_0\,r^{4/(k+1)} + \ldots +
g_{2{q}}\,r^{(4q+4)/(k+1)} \right ] \,
 \label{gegeSExt}
 \ee
with a fixed $k$ (say, for the sake of simplicity, $k=0$) and
growing integers $q$, an efficient use of the above large$-D$
approach does not seem too promising.

In an alternative semi-analytic approach to the problem
(\ref{rad}) + (\ref{gegeSExt}) one could try to employ a
power-series method where
 \be
{ \psi^{(PS)}_{(q,k)}}(r)
 \approx r^{\ell+1} \times U(r)
 \times \exp W(r)\,,
 \label{neMagan}
 \ee
and where approximations $U(r) = polynomial\,[r^{2/(k+1)}]$ (of a
sufficiently large degree $m_U$) and $W(r) =
polynomial\,[r^{2/(k+1)}]$ (of any degree $m_W$) may be
re-constructed in a more or less algebraic manner after an
insertion of this ansatz in the original differential equation.
Unfortunately, virtually all practical implementations of this
method prove even less efficient in computations \cite{ZnojilHD}.

The third eligible non-numerical approach to the construction
represents a modification of the power-series method where also
the second polynomial degree $m_U< \infty$ is fixed. This method
has been proposed by E. Magyari \cite{Magyari} and is based on the
evaluation of the solutions (\ref{neMagan}) via an {\it ad hoc}
tuning of the potential (\ref{gegeSExt}) itself. Basically, the
recipe requires the absence of any errors in eq. (\ref{neMagan}).
This idea just extends the well known polynomial solvability of
eq. (\ref{rad}) + (\ref{gegeSExt}) at $q=k=0$ (harmonic
oscillator) and $q=k-1=0$ (Coulomb field). The details of the
Magyari's recipe have been reviewed in ref. \cite{PI} where we
emphasized that its {\em practical} merits are virtually
non-existent because the underlying and obvious selfconsistency
requirements (we called them there the Magyari-Schr\"{o}dinger
equations) may be characterized as a coupled set of determinantal
equations which are extremely complicated to solve in general.

The main reason why we decided to pay attention to the Magyari's
approach is that we discovered, many year ago \cite{Dubna}, that
this method encounters enormous simplifications during a limiting
transition to the large dimensions $D\gg 1$. In this setting one
has to follow very consequently all the analogies with the
harmonic oscillators. Firstly, one assumes that the polynomial
$W(r)$ in the exponential represents {\em an exact} $r \gg 1$
asymptotic solution at any $q$ and $k$. Both the degree $m_W$ and
all the coefficients in $W(r)$ are uniquely determined, in this
way, by our choice of the potential (for example, we have $W(r) =
-r^2/2$ for harmonic oscillator). On this background, the
Magyari's {\em second } key idea parallels the essence of the
quasi-exact solvability (i.e., an incomplete solvability, see
\cite{Ushveridze}) and extends, to all the $q\geq 1$ cases, the
requirement that {\em our} choice of the polynomial $U(r)$ of {\em
any} integer degree $N = m_U$ leads to  {\em some} exact solution
(\ref{neMagan}) for a suitably {modified, adapted} polynomial
potential~(\ref{gegeSExt}).

Our present text is inspired by the series of demonstrations that
the latter combined method proves unexpectedly efficient in
practice. As we mentioned, our study was initiated by the
observation that for one of the most popular, viz., quartic
polynomial interaction, the complicated Magyari-Schr\"{o}dinger
equations exhibit a remarkable and fairly surprizing
simplification. In the next stage of our work on this project
\cite{jmch} we choose the ``first nontrivial" $q=2$ oscillators
and complemented the above observations by the formulation of the
method of an explicit perturbative construction of $1/D$
corrections. We must emphasize that the improved convergence of
our innovated solutions (which were defined by the series in the
powers of $1/D^2$) was in a sharp contrast to the steady worsening
of the performance of the above-mentioned large$-D$ expansions at
the larger $q$. Moreover, in the ``minimal" sextic example at
$q=1$, we succeeded in showing that and why our innovated series
in the powers of $1/R \approx 1/D^{1/4}$ was absolutely convergent
\cite{Drumi}.

A real climax of our effort came with the papers \cite{PI} and
\cite{copnt} where we performed an explicit construction of the
zero-order solutions and discovered that at the next few less
trivial exponents $q=3$, $q=4$ and $q=5$ we were still able to
construct the $D \to \infty$ solutions in closed form.
Unfortunately, we were just able to work by the brute-force
methods, based on the direct solution of the $D \to \infty$ limit
of the Magyari-Schr\"{o}dinger coupled algebraic nonlinear
equations by the elimination method using the Groebner bases
\cite{Passau}. For this reason, we were never able to find any
solution at $q \geq 6$. This was also the main motivation for the
study which we are going to describe in what follows.

\section{A concise formulation of the problem  \label{Mag}}

\subsection{Asymptotic Magyari-Schr\"{o}dinger equations}

In a way discussed thoroughly in \cite{PI}, the use of our ansatzs
in the limit $D \to \infty$ implies an immediate reduction of the
{\em differential} radial Schr\"{o}dinger equation to its
algebraic equivalent
  \begin{equation}
 \label{trap}
 \left(  \begin{array}{cccccc}
 s_1 & 1 & & & & \\
 s_2 & s_1 & 2 & &  & \\
 \vdots&  &\ddots & \ddots & & \\
 s_q & \vdots &  &s_1 & N-2& \\
 N-1& s_q & &  & s_{1} & N-1 \\
 &N-2& s_q & & \vdots& s_{1}  \\
 &&\ddots&\ddots&&\vdots\\
 &&& 2 & s_q&  s_{q-1}   \\ &&&& 1 & s_q
\end{array} \right)
 \left(  \begin{array}{c}
 {p}_0\\
 {p}_1\\
\vdots \\
 {p}_{N-2}\\
 {p}_{N-1}
\end{array} \right)
=
0\,.
 \end{equation}
Here we are going to study its solvability, referring to \cite{PI}
(we shall call it paper PI in what follows) for all the
explanations of its origin and interpretations. Calling this
deeply nonlinear non-square-matrix algebraic problem the
Magyari-Schr\"{o}dinger equation, we may just summarize that the
$N$ quantities $p_j$ are in a one-to-one correspondence with the
Taylor coefficients in the parts $U(r)$ of wavefunctions
(\ref{neMagan}). Moreover, all $q$ generalized eigenvalues $s_a$
[cf. their definitions $s_a=s_a(g_{a-2})$ by eq. (30) in PI] are
just certain re-scaled forms of the energies and/or of coupling
constants in our potential (\ref{gegeSExt}).

\subsection{Upside-down symmetry}

As we already explained in paper PI, the practical use of the
explicit quasi-exact (QE) solutions requires a purely numerical
determination of their QE-compatible eigenvalues $g_{a-2}$ with
$a=1,2,\ldots,q$.  This means that the simplified
Magyari-Schr\"{o}dinger eq.  (\ref{trap}) should be interpreted as
an implicit definition of a QE solution {\em on a pre-determined
level of a finite precision}.  All the sufficiently small ${\cal
O}(1/D)$ variations of the parameters should be ignored, within
this convention, as negligible and irrelevant.  Alternatively,
they may all be treated, if necessary, by perturbation methods in
a way exemplified in \cite{KyjevI} at $q=1$ and in \cite{jmch} at
$q=2$.

In such a setting, one of the main messages delivered by paper
PI was that all the perturbation constructions of any type may
remain exact and non-numerical, at the first few smallest $q$ at
least.  This follows from the observation that all the exact QE
parameters in PI proved expressible through integers in the
zero-order limit $D \to \infty$.  The necessary condition of
this simplification lies in the elementary form of our
algebraized QE Magyari-Schr\"{o}dinger equation (\ref{trap}).
Still, one has to overcome a few further obstacles. In
particular, the brute-force origin of the results in PI made it
impossible to move beyond $q\leq 5$.  In the other words, the
key weakness of paper PI lies in the too rapid growth of the
difficulties with the increasing $q$.  One needs all the
capacity of the available computers to reveal the structure of
solutions at the first few $q$.  This means that whenever one
needs an improvement of the insight in the structure of the
solutions, one has to make a better use of their symmetries.

In our present continuation and completion of paper PI we are
going to exploit
the most obvious symmetry of eq. (\ref{trap}) with respect to
its upside-down transposition. For this purpose we may
modify our notation slightly, replacing eq. (\ref{trap}) by its
reincarnation
  \begin{equation}
 \label{trapava}
 \left(  \begin{array}{cccccc}
 \alpha_1 & 1 & & & & \\
 {\alpha}_2 & {\alpha}_1 & 2 & &  & \\
 \vdots&  &\ddots & \ddots & & \\
  \tilde{\alpha}_1 & \vdots &  &{\alpha}_1 & N-2& \\
 N-1&  \tilde{\alpha}_1 & &  & {\alpha}_{1} & N-1 \\
 &N-2&  \tilde{\alpha}_1 & & \vdots& {\alpha}_{1}  \\
 &&\ddots&\ddots&&\vdots\\
 &&& 2 &  \tilde{\alpha}_1&   \tilde{\alpha}_{2}   \\ &&&& 1 & \tilde{\alpha}_1
\end{array} \right)
 \left(  \begin{array}{c}
 {p}_0\\
 {p}_1\\
\vdots \\
  \tilde{p}_{1}\\
  \tilde{p}_{0}
\end{array} \right)
=
0\,.
 \end{equation}
However trivial, such a change of notation implies that all the
separate lines have a tilded partner in this set. For example,
the definition ${p}_0= - {p}_1/\alpha$ (where $\alpha=\alpha_1$
is tacitly assumed non-zero) is {\em always} accompanied by its
tilded counterpart $\tilde{p}_0= -\tilde{p}_1/\tilde\alpha$
whenever $\tilde\alpha=\tilde\alpha_1\neq 0$, etc.

\section{Solutions of the symmetrized recurrences}

\subsection{Trivial case at $N=2$}

In our tilded notation, the first nontrivial version of our problem
(\ref{trapava}) with $N=2$ is particularly transparent and instructive,
  \begin{equation}
 \label{trap2b}
 \left(  \begin{array}{cc}
 \alpha_1 & 1 \\
 \alpha_2 & \alpha_1\\
 \alpha_3 & \alpha_2\\
 \vdots&  \vdots\\
 \tilde{\alpha}_1&  \tilde{\alpha_2}   \\
 1 & \tilde{\alpha}_1
\end{array} \right)
 \left(  \begin{array}{c}
 {p}\\
 \tilde{p}
\end{array} \right)
=
0\ .
 \end{equation}
Its upside-down or ``tilding" symmetry separates its first and last
line as  giving a constraint upon the doublet of unknowns
$\alpha_1=s_1=\alpha$ and $\tilde{\alpha}_1=s_q=\tilde{\alpha}$
in a tilding-symmetric manner,
  \begin{equation}
 \label{trap2bdet}
 \det \left(  \begin{array}{cc}
 \alpha & 1 \\
 1 & \tilde{\alpha}
\end{array} \right) = 0, \ \ \ \ \ \ \ \ \ \ \ \ \
\alpha\,\tilde{\alpha}=1\,.
 \end{equation}
The fully tilding-symmetric way of dealing with the rest of eq.
(\ref{trap2b}) consists now in the equivalence of
its recurrent downward or upward
treatment.

After we normalize $p_1=\tilde{p}=1$, the first line defines
$p_0={p}=-1/\alpha$ and may be omitted. Step-by-step, the
$k-$th line of eq. (\ref{trap2b}) may be multiplied by the factor
$\tilde{\alpha}/\alpha^{k-1}$. This converts the old components
into the known constants and we get the following sequence of
definitions
 \be
 \tilde{\alpha}\alpha=1, \ \ \
 \tilde{\alpha}\alpha_2/\alpha=1, \ \ \
 \tilde{\alpha}\alpha_3/\alpha^2=1, \ldots\,,
 \ee
with elementary consequence: $\alpha_k=\alpha^k$. We may imagine that
the last definition prescribes that $\alpha_{q+1}=\alpha^{q+1}=1$.
This equation may be read as a boundary condition for our
recurrences, fixing the physical value of our single free parameter
$\alpha$. It has  many unphysical
complex roots and just a single {\em real} one,
viz., the physical root $\alpha=1$
at any even $q$. Similarly, {\em two}
 different real roots $\alpha=\pm 1$ become available at
all the odd $q$'s. This makes the final reconstruction of all the
original QE-compatiblity ``eigenvalues" $s_1, \ldots, s_q$
trivial.

We may add a comment. Knowing that the last line of recurrences
(\ref{trap2b}) defines the function of $\alpha$
($\alpha_{q+1}=\alpha^{q+1}$) with a prescribed value
($\alpha^{q+1}=1$), the latter constraint may be interpreted  as
an algebraic equation which fixes the eligible values of $\alpha$.
Such a type of the boundary condition is not unique. The same role
may be played by any other line of eq. (\ref{trap2b}), once we
re-direct these recurrences and demand that
 \be
 \tilde{\alpha}_{1+j}=\alpha_{q-j}
 \label{rela}
 \ee
at any shift $j \leq q-1$. At $N=2$ all this is trivial since after
being multiplied by $\alpha^{j+1}$, all relations (\ref{rela})
degenerate to the same rule $\alpha^{q+1}=1$.

\subsection{A transition to the single variable at $N=3$}

A separate treatment of the first nontrivial
$N=3$ version of eq. (\ref{trap})
is necessary in the degenerate case with $p_1 =
0$. We may infer that $s_1=s_3=s_5 = \ldots =s_q= 0$.
This means that $q=2Q+1$ must be odd and that we in effect return to
the $N=2$ structure. We only have to replace the old unknowns $s_k$
by the new ones, re-scaled to $s_{2k}/2$. Otherwise, the construction
of the solutions remains strictly the same, giving the nontrivial
roots $s_{2k}=2\,\varrho^k$ where $\varrho^{Q+1}=1$.

In what follows, similar detailed qualification will be omitted and,
with the degenerate solutions ignored, we shall normalize
$p_1=1$ at $N=3$, etc.

From the two outer lines
of the $N=3$ version of eq. (\ref{trap})
we deduce that $p_0=-1/\alpha$ while
$p_2=-1/\tilde{\alpha}$. The rest of equation (\ref{trap}) acquires
the tilding-symmetric matrix form
  \begin{equation}
 \label{trapgen}
 \left(  \begin{array}{ccc}
 \beta & \alpha & 2 \\
\gamma& \beta & \alpha \\ \delta& \gamma& \beta \\
 \vdots& \vdots & \vdots  \\
  \tilde{\alpha} & \tilde{\beta}& \tilde{\gamma}\\
 2 &  \tilde{\alpha} & \tilde{\beta}
\end{array} \right)
 \left(  \begin{array}{c}
-1/\alpha\\
 1\\
-1/\tilde{\alpha}
\end{array} \right)
=
0\ .
 \end{equation}
It gets facilitated when pre-multiplied by an auxiliary row vector.
This observation results from the step-by-step analysis of this
system of equations re-written in the form
 \be
 {\tilde{\alpha}}\,\beta/{\alpha}=\tilde{\alpha}\alpha - 2,\ \ \
 \
 {\tilde{\alpha}}\,\gamma/{\alpha}=\tilde{\alpha}\beta -\alpha,\ \ \ \
 {\tilde{\alpha}}\,\delta/{\alpha}=\tilde{\alpha}\gamma -\beta,\
 \ \ \
 \ldots\,.
 \label{system}
 \ee
In the first item the right-hand-side part $\tilde{\alpha}\alpha
-2=\alpha \tilde{\alpha}-2\equiv \xi-2= \tilde{\xi}-2$ is tilding
symmetric.  This means that the same tilding-invariance must hold for
the left-hand-side expression as well. The second item is not
tilding-invariant but the invariance is restored after we divide all
this equation by $\alpha$. This gives a consistent picture because
one can deduce that also in all the subsequent rows the full
tilding-invariance is achieved when we replace $\alpha$, $\beta$,
$\gamma$, $\ldots$ by their renormalized and tilding-invariant forms
$\alpha\tilde{\alpha}/\alpha^0$, $\beta\tilde{\alpha}/\alpha$,
$\gamma\tilde{\alpha}/\alpha^2$, $\ldots$, respectively. In the other
words, the system (\ref{system}) must be pre-multiplied by the row of
the factors $1,\,1/\alpha,\,\tilde{\alpha}
/\alpha,\,\tilde{\alpha}/\alpha^2 ,\,\tilde{\alpha}^2/\alpha^2
,\,\ldots$ obtained, in recurrent manner, by the multiplication by
the quotient which depends on the parity, i.e., equals to $1/\alpha$
and to $\tilde{\alpha}$ in subsequent steps. This means that the even
and odd items in eq. (\ref{system}) have a different structure.

This difference may be reflected by the change of the notation. Once
we put $\alpha=s_1=a$, $\beta=s_2=A$, $\gamma=s_3=b$, $\delta=s_4=B$,
$\epsilon=s_5=c$ (while denoting also $\tilde{\alpha}=s_q=\tilde{a}$
etc) etc, equation (\ref{trap}) acquires another formally
tilding-symmetric matrix form
  \begin{equation}
 \label{trapgenmod}
 \left(  \begin{array}{ccc}
 a & 1 & 0 \\
 A & a & 2 \\
b& A & a \\
B& b& A \\
 \vdots& \vdots & \vdots
\end{array} \right)
 \left(  \begin{array}{c}
\tilde{a}/a\\
 -\tilde{a}\\
 1
\end{array} \right)
=
0\ .
 \end{equation}
The pair of the old variables $a$ and $\tilde{a}$ must be replaced by
their tilding-invariant product $\xi = a\,\tilde{a} = \tilde{\xi}$
and its tilding-covariant complement $\rho = a/\tilde{a} =
1/\tilde{\rho}$. In an opposite direction, whenever needed, we may
re-construct $a$ and $\tilde{a}$ from the two quadratic relations
$a^2=\rho\,\xi$ and $\tilde{a}^2 = \xi/\rho$, i.e., up to an
inessential
indeterminacy in sign. After we abbreviate
  \ben
 \Sigma_1=\frac{A}{\rho}, \ \ \
 \Sigma_2=\frac{B}{\rho^2}, \ \ \
 \Sigma_3=\frac{C}{\rho^3}, \ \ \ \ldots\,, \ \ \
 \sigma_{1}=
 \frac{1}{a}\, \frac{a}{\rho^0},\ \ \
 \sigma_{2}=
 \frac{1}{a}\, \frac{b}{\rho},\ \ \
 \sigma_{3}=
 \frac{1}{a}\, \frac{c}{\rho^2},\ \   \ldots
 \een
and postulate that $\Sigma_{0}=2$ and  $\sigma_{1}=1$, this procedure
results in the conclusion that our recurrences may be re-written as
the following sequence of the coupled pairs of the recurrent
relations,
  \be
 \label{reku}
 \Sigma_k=\xi\,\sigma_k-\Sigma_{k-1}\,,\ \
 \ \ \ \ \
 \sigma_{k+1}=\Sigma_k-\sigma_{k},\ \
 \ \ \ \ \ \ \ \ \ k = 1, 2, \ldots\ .
  \ee
One re-interprets eqs. (\ref{reku}) as the mere recurrent definition
of the auxiliary sequence of functions of our auxiliary real
variable~$\xi$,
 \ben
\Sigma_1=\xi-2, \ \ \sigma_2=\xi-3, \ \ \Sigma_2=\xi^2-4\,\xi+2, \ \
\ldots\ .
  \een
We see that the functions  $\Sigma_k(\xi)$ and $\sigma_{k+1} (\xi)$
are, by their construction, both polynomials of the same degree~$k$.

Our final change in the notation will prescribe $\xi$ replaced by
$\xi=4x^2$, with $\Sigma_k$ represented as
$\Sigma_k=2\,T_{2k}(x)$ and with $\sigma_k$ re-scaled into
$\sigma_k=\,T_{2k-1}(x)/x$. We notice that our recurrences
become simpler in the new notation but what is more important is
that after such a transformation, our new polynomials $T_n(x)$
{\em coincide precisely} with the classical orthogonal
Chebyshev polynomials of the first kind \cite{Ryzhik},
 \be
 T_0(x)=1, \ T_1(x)=x, \ T_2(x)=2x^2-1, \ \ldots\,.
 \ee
In this sense, our $N=3$ recurrences are {\em solved exactly} in
closed form.

\subsection{The two tilding-covariant variables at $N=4$}

Using the abbreviations $\alpha=s_1$ and $\tilde{\alpha}=s_q$
etc., let us interpret the $N=4$ QE recurrences
(\ref{trap}) as a tilding-symmetric problem
 \begin{equation}
\label{trap4} \left(
\begin{array}{cccccc}
\alpha &1&&  \\
\beta &\alpha &2&  \\
 \gamma&  \beta &\alpha &3  \\
 \delta& \gamma&  \beta &\alpha  \\
 \epsilon& \delta& \gamma&  \beta   \\
 \vdots  &\vdots & \vdots & \vdots \\
\\ 3 & \tilde{\alpha} & \tilde{\beta} & \tilde{\gamma}
\\&2 & \tilde{\alpha}& \tilde{\beta}
\\&&1 & \tilde{\alpha}
\end{array} \right)
 \left(  \begin{array}{c}
p\\t\\
\tilde{t}\\
\tilde{p}
\end{array} \right)
=
0\,.
 \end{equation}
The first and the last line may be dropped as defining merely
$p=-t/\alpha$ and $\tilde{p}=-\tilde{t}/\tilde{\alpha}$. From the
next two outer lines we may express $t$ in terms of $\tilde{t}$ or
{\it vice versa}. Once we normalize $\tilde{t}=1$ and re-parametrize
$s_2=\beta=\beta(b)=\alpha^2+2\,b\,\alpha$ and, in parallel,
$s_{q-1}=\tilde{\beta}=\tilde{\beta}(\tilde{b})=\tilde{\alpha}^2+
2\,\tilde{b}\,\tilde{\alpha}$, the vanishing of the
related secular two-by-two
determinant may be re-read as the constraint
$b\,\tilde{b} = 1$. The rest of eq. (\ref{trap4}) reads
  \begin{equation}
 \label{trap4b}
 \left(  \begin{array}{cccccc}
 \gamma\mbox{}&  \beta(b) &\alpha &3  \\
 \delta& \gamma&  \beta(b) &\alpha  \\
 \epsilon& \delta& \gamma&  \beta(b)   \\
 \vdots  &\vdots & \vdots & \vdots \\
\\ 3 & \tilde{\alpha}
 & \tilde{\beta}(\tilde{b})
 & \tilde{\gamma}
\end{array} \right)
 \left(  \begin{array}{c}
 -\tilde{b}\,\tilde{\alpha}/\alpha\\
 \tilde{b}\,\tilde{\alpha}\\
\tilde{\alpha}\\
-1
\end{array} \right)
=
0\,.
 \end{equation}
Its first row expresses $\gamma$ as a function of $\alpha$ and $b$
and $\tilde{\alpha}$ and $\tilde{b}$,
 \be
 \gamma\,\tilde{b}\,\tilde{\alpha}/\alpha=Q=
\beta(b)\,\tilde{b}\,\tilde{\alpha}+ \alpha\,\tilde{\alpha}-3\,
.
 \ee
Fortunately, the quantity $Q$ may be read as a polynomial in the
mere two new, auxiliary variables $Z=\alpha\,\tilde{\alpha}
\equiv \tilde{Z}$ and $Y = \alpha\,\tilde{b}= Z/\tilde{Y}$.
In the light of the preceding subsection, this function $Q =
Q(Z,Y)=Z\,Y+3\,Z-3$ of two variables will obviously play the
role of a generalized Hermite polynomial.  We may only regret
that it is not a tilding-invariant function anymore.

The second row of eq. (\ref{trap4b})
must be multiplied by $Y\,\tilde{\alpha}$ to give
the next-order polynomial in the same two variables,
 \be
R = \delta \tilde{b}\,\tilde{\alpha}^2\,Y/\alpha=
 Z\,Y\,Q(Z,Y) + Z^2\,Y + 2\,Z^2-Z\,Y=R(Z,Y)\,.
 \ee
This defines the new quantity (= a rescaled $\delta$) and we may
proceed to the third row multiplied by $Y\,\tilde{\alpha}^2$,
 \be
 S=
 \epsilon \tilde{b}\,\tilde{\alpha}^3\,Y/\alpha=
 Z\,R(Z,Y)
 + Z^2\,Q(Z,Y)- Z^2\,Y - 2\,Z^2=S(Z,Y)\,.
 \ee
After the multiplication by  $Y^2\,\tilde{\alpha}^3$ the
fourth row reads
 \be
 T=
 \zeta \tilde{b}\,\tilde{\alpha}^4\,Y^2/\alpha=
  Z\,Y\,S(Z,Y)+Z^2\,R(Z,Y)-Y\,Z^2\,Q(Z,Y)\,
 \ee
with the next factor $Y^2\,\tilde{\alpha}^4$
giving the next, fifth row
 \be
 U=
 \eta \tilde{b}\,\tilde{\alpha}^5\,Y^2/\alpha=
 Z\,T(Z,Y)+ Z^2\,S(Z,Y)-Z^2\,R(Z,Y)
 \ee
etc.
Step by step we construct, in this manner, the two sequences of
functions denoted as $P_n(Y,Z)$ and $Q_n(Y,Z)$
and defined by the pair of the
coupled recurrences,
\be
\ba
P_{n+1} = Y\,Z\,Q_n + Z^2\,P_n-
 Y\,Z^2\,Q_{n-1},\\
Q_{n+1} = Z\,P_{n+1} + Z^2\,Q_n-
 Z^2\,P_n,\\
\ \ \ \ \ \ \ \ \ \ \ \ \ \ \ \
n = 0, 1, \ldots
\ea
\ee
from the initial values $Q_{-1}=1/Z$, $P_0=Y+2$ and
$Q_0=Q=Y\,Z+3\,Z-3$ generating $R=P_1$ etc.

In a way paralleling the previous $N=3$ case, we might slightly
modify the functions and define $P_{n+1}=\sqrt{Y}\,W_{2n+1}$
while $Q_{n+1}=W_{2n+2}$. It is easy to verify that we can now
use just the single common recurrence
\be
\ba
W_{n+1} = \sqrt{Y}\,Z\,W_n + Z^2\,W_{n-1}-
 \sqrt{Y}\,Z^2\,W_{n-2}\,,\\
\ \ \ \ \ \ \ \ \ \ \ \ \ \ \ \
n = 0, 1, \ldots
\ea
\ee
with the merely slightly modified initialization by
$W_{-2}=Q_{-1}=1/Z$, $W_{-1} =P_0/\sqrt{Y}=\sqrt{Y}+2/\sqrt{Y}$
and $W_0=Q_0=Q=Y\,Z+3\,Z-3$.
We may see
the clear parallels with the previous $N=3$ case,
noticing
that the polynomials $W_{3n}$
and $W_{3n-1}$ are both divisible by $Z^{2n}$ while
 $W_{3n-2}$ is only divisible by $Z^{2n-1}$.
We shall skip the further technical analyses of this sort here.

\newpage

\section{Matching and secular polynomials at $N=3$}

For the sake of simplicity, let us only pay attention to the choice
of $N=3$. Then, the knowledge of the closed form of the polynomials
$\Sigma_k(\xi)$ and $\sigma_{k+1}(\xi)$ enables us to define the
explicit values of {\em all} our coupling constants as functions of
the mere {\em two} parameters $a$ and $\tilde{a}$ entering
$\xi=a\,\tilde{a}$ and $\rho=a/\tilde{a}$,
 \ben
 a_1=a=a\,\rho^0\,\sigma_1(\xi), \ \ \
 A_1=A=\rho\,\Sigma_1(\xi), \ \ \ a_2=b=a\,\rho\,\sigma_2(\xi), \ \
 \een
 \be
A_2=B=\rho^2\,\Sigma_2(\xi), \ \
a_3= c=a\,\rho^2\,\sigma_3(\xi), \ \ \ldots \
. \label{definice}
  \ee
One could also have constructed this general solution of our
recurrences (\ref{trapgen}) in an opposite, upward direction. For
this purpose, it suffices when all the above formulae are modified by
a consequent application of the tilding operation.

We have seen that the $N=3$ case operates with two unknowns.
At the same time,
the set of recurrences (\ref{trapgen}) contains
precisely two redundant
items. In one extreme example we may read whole this set as a
sequence of definitions of $\beta=s_2=s_2(s_1,s_q)$,
$\gamma=s_3=s_3(s_1,s_q)$, \ldots, $s_{q+1}=s_{q+1}(s_1,s_q)$
where the last two lines are redundant since we already knew
the outcome, viz., $s_{q+1}=2$ and $s_{q}(s_1,s_q)=s_q$.  This
may be understood as a source of our final pair of boundary
conditions determining the QE-compatible values of the pair of
the unknown parameters.

In a way paralleling the previous $N=2$ example, any other two
lines of eq. (\ref{trapgen}) might be selected as boundary
conditions.  In contrast to the $N=2$ example, almost all of the
non-extreme choices of matching conditions would be preferable
in practice, lowering the degree of the resulting secular
polynomials.

This observation deserves to be explained in more detail.
Indeed, it makes sense to distinguish between the four possible
selections of the optimal matching conditions.

\subsection{$q=4K$}

Whenever $q=4K$ where $K = 1, 2, \ldots$, the above-mentioned
recurrent construction may be started, simultaneously, at both the
upper and lower end of eq. (\ref{trapgen}). Without any difficulties
and using eq. (\ref{definice}), the recipe defines {\em all} the
unknown quantities, i.e., the doublets of pairs of the values
 \be
 \left (
 a_j, \ A_j
 \right ), \ \
 \left (
 \tilde{a}_j, \ \tilde{A}_j
 \right ), \ \ \ \ \ \ j = 1, 2, \ldots , K\,.
 \label{setuno}
 \ee
The two middle lines of eq.  (\ref{trapgen}) define the other
two redundant functions (or ``non-existent couplings") $a_{K+1}$
and $\tilde{a}_{K+1}$. This induces no real difficulty since the
two parameters $a_1=a$ and $\tilde{a}_1=\tilde{a}$
are not yet specified. The latter two definitions are not
redundant, therefore, as they have to fix these initial
values.

The inspection of eq. (\ref{trapgen}) reveals that our symbol
$a_{K+1}$ is an alternative name for another and well defined
coupling $\tilde{A}_K$. Similarly, the quantity
$\tilde{a}_{K+1}$ is an ``alias" for $A_K$.  We
determine the missing QE roots $a$ and $\tilde{a}$ via the
two redundant equations $a_{K+1}=\tilde{A}_K$ and
$\tilde{a}_{K+1}=A_K$ or, in the notation of eq.
(\ref{definice}),
 \ben
 a\,\sigma_{K+1}\,{\rho^K}=
 \Sigma_K\,
 \tilde{\rho}^K\,, \ \ \ \ \ \ \ \
 \ \ \ \ \ \
 \tilde{a}
 \,\sigma_{K+1}\,\tilde{\rho}^K=
 \Sigma_K\,{\rho^K}\,.
 \een
Their ratio reads $\rho^{4K+1}=1$ and gives the unique real root
$\rho = 1$. Our first conclusion is that we must have
$a=\tilde{a}$.  The above two equations
coincide and any of them represents our ultimate matching
condition or constraint imposed upon $\xi=a^2$,
 \be
 a\,\sigma_{K+1}(a^2)= \Sigma_K(a^2)\,,
 \ \ \ \ \ \ \ \ \ \ \ \ q = 4K\ .
 \ee
A sample of its roots may be found in Table~1 at $q=4$ and
$q=8$. The inspection of the subsequent Table 2 reveals that
with the further growth of $K$, the determination of these roots
becomes purely numerical very quickly.

\subsection{$q=4K+2$}

After a move to $q=4K+2$ with $K = (0,)\, 1,2,  \ldots$, the previous
recipe does not change too much. This time we define all the unknowns
in a reversed order,
 \be
 \left (
  A_j, \ a_{j+1}
 \right ), \ \
 \left (
 \tilde{A}_j, \ \tilde{a}_{j+1}
 \right ), \ \ \ \ \ \ j = 1, 2, \ldots , K\,.
 \label{setdva}
 \ee
{\it Mutatis mutandis} we find that the central part of eq.
(\ref{trapgen}) defines the other ``non-existent" couplings
$A_{K+1}$ and $\tilde{A}_{K+1}$ so that the  doublet of
equations $A_{K+1}=\tilde{a}_{K+1}$ and
$\tilde{A}_{K+1}=a_{K+1}$ leads to another set of the
selfconsistency conditions,
 \ben
 a\,\sigma_{K+1}\,{\rho^K}=
 \Sigma_{K+1}\,
 \tilde{\rho}^{K+1}\,, \ \ \ \ \ \ \ \
 \ \ \ \ \ \
 \tilde{a}
 \,\sigma_{K+1}\,\tilde{\rho}^K=
 \Sigma_{K+1}\,{\rho^{K+1}}\,.
 \een
Their ratio degenerates to the modified constraint
$\rho^{4K+3}=1$ with the same unique real root as above,
$a/\tilde{a}=\rho = 1$. Both our innovated identities
coincide,
 \be
 a\,\sigma_{K+1}(a^2)= \Sigma_{K+1}(a^2)\,,
 \ \ \ \ \ \ \ \ \ \ \ \ q = 4K+2\,
\label{secula}
 \ee
and guarantee the desired matching. Their numerical aspects are
sampled again in Table~1 (easily solvable cases at $K=0,1$ and
$2$). The adjacent Table~2 complements this list and facilitates
the determination of the explicit form of the secular equation
(\ref{secula}) at all the integers $K$.

\subsection{$q=4K+1$}

The subset of odd $q=4K+1$ with $K = (0,)\, 1,2, \ldots$
requires a more careful analysis.  Although we  have the same
complete list (\ref{setdva}) of the definitions of the QE-fixed
couplings as above, its {last two} items are defined twice, in
two different ways. Their necessary compatibility represented by
the relation $a_{K+1}=\tilde{a}_{K+1}$ or rather
 \ben
 a\,\sigma_{K+1}\,{\rho^K}=
 \tilde{a}
 \,\sigma_{K+1}\,\tilde{\rho}^K\,
 \een
implies that $\rho^{2K+1}=1$ so that we must put $a/\tilde{a}=\rho
= 1$. In the light of this conclusion, the other two
consequences $A_{K+1}=\tilde{A}_{K}$ and $\tilde{A}_{K+1}=A_{K}$ of
the two other next-to-central rows of eq.
(\ref{trapgen}) coincide and give the same ultimate matching rule
 \be
 \Sigma_{K+1}(\xi)= \Sigma_{K}(\xi)\,,
 \ \ \ \ \ \ \ \ \ \ \ \ q = 4K+1\,.
 \ee
Its numerical performance appears illustrated by the
corresponding subset of roots in Table~3.


Marginally, let us note that
for the specific exponents $q=4K+1$,
the secular polynomial
may be re-written in the compact form
 \ben
R^{(K,-)}(\xi) =
(\xi-4)\,
\left [
\left (
\ba
2K\\0
\ea
\right )\,\xi^K
-
\left ( \ba 2K-1\\1 \ea \right )\,\xi^{K-1} + \left ( \ba 2K-2\\2 \ea
\right )\,\xi^{K-2} +\ldots \right .
 \een
 \be
 \left .
 \ldots
(-1)^{K+2} \left ( \ba K+2\\ K-2 \ea \right )\,\xi^2 + (-1)^{K+1}
\left ( \ba K+1\\ K-1 \ea \right )\,\xi + (-1)^{K} \left ( \ba K\\
K
\ea
\right ) \right ]\,.
\ee
The
secular polynomials $\sum_{m=0}^K\,\xi^m\,d_m^{[K]}$
contain the $(K+1)-$plets of coefficients
${\cal K}^{(K)}=
\left (d_K^{[K]},\,
d_{K-1}^{[K]},\ldots,
d_0^{[K]} \right )$
such that
${\cal K}^{(0)}=(1)$,
${\cal K}^{(1)}=(1,-1)$,
${\cal K}^{(2)}=(1,-3,1)$,
${\cal K}^{(3)}=(1,-5,6,1)$,
etc.
This rule parallels the even$-q$ recipe of Table~2.

\subsection{$q=4K+3$}

The last possible choice of the odd exponents $q=4K+3$
(with $K = (0,)\, 1,2, \ldots$) in the
potentials $V_{(q,n)}(r)$ of eq. {\ref{gegeSExt})
leads to
a routine completion of all the above analysis. A marginal
modification of the list (\ref{setuno}) is needed to specify all
the necessary QE couplings, recurrently determined as functions
of $a$ and $\tilde{a}$ only,
 \be
 \left (
 a_j, \ A_j
 \right ), \ \
 \left (
 \tilde{a}_j, \ \tilde{A}_j
 \right ), \ \ \ \ \ \ j = 1, 2, \ldots , K+1\,.
 \label{setunodva}
 \ee
Nevertheless, the results of the matching become slightly
different this time. Although the first, central-line rule
$A_{K+1}=\tilde{A}_{K+1}$ prescribes merely
 \ben
 \Sigma_{K+1}\,{\rho^{K+1}}=
 \Sigma_{K+1}\,\tilde{\rho}^{K+1}\,
 \een
its consequence $\rho^{2K+2}=1$ admits the two alternative signs
in the resulting $a/\tilde{a}=\rho = \pm 1$. Under this condition,
the other two equations (in detail,
$a_{K+2}=\tilde{a}_{K+1}$ and its tilding-conjugate
$\tilde{a}_{K+2}=a_{K+1}$) coincide as well, giving the same final
condition
 \be
 \sigma_{K+2}(\xi)= \sigma_{K+1}(\xi)\,,
 \ \ \ \ \ \ \ \ \ \ \ \ q = 4K+3\,.
 \ee
Curiously enough, this equation is the most easily solvable
implicit definition of the QE roots $\xi = a^2=\tilde{a}^2$
(cf. Table~4).

Even the shortest glimpse at the results of the factorization of
the effective secular polynomial $R^{(K,+)}(\xi) =
\sigma_{K+2}(\xi)- \sigma_{K+1}(\xi)$ reveals that the sequence
of the exponents $q = 4K+3$ might be viewed as the most
privileged one.  The search for its QE roots becomes by far the
easiest.  After we omit the roots $\xi = 0$ and
$\xi = 4$ as trivial, we encounter another unexpected
and purely empirically observed
symmetry. Indeed, the secular roots $a_1=\sqrt{\xi_{\pm
n}}=\sqrt{ 2 \pm \Xi_n^{[K]}}$
listed in Table~4
 at the indices $q=4K+3$ appear to be
of the very similar form.
It is realy instructive to list a few sample
{\rm distances } $\Xi_n^{[K]}$
{\rm of }
$a^2_1$
  from their median = 2. We have
 $\Xi_n^{[1]} = 0$,
 $\Xi_n^{[2]}=1$,
 $\Xi_n^{[3]}=0$ or $\sqrt{2}$,
two values of $\Xi_n^{[4]}= (\sqrt{5}\pm 1)/2$,
the three values of
 $\Xi_n^{[5]}
 = 0, \ 1$ and $\sqrt{3}$.
  One may see
 that the full set of the secular roots $a_1=\sqrt{\xi}$
exhibits a weird regularity manifested by a reflection symmetry
with respect to the center at $\xi_c=2$. All roots become
tractable as certain quasi-conjugate pairs $\xi=\xi_{\pm n}={ 2
\pm \Xi_n^{[K]}}$.
In this sense, the results listed in
Table 4 may be tentatively extrapolated to all the values of $q$.
Indeed,
once we omit the permanent pair of the
 minimal and maximal
QE-compatiblity roots $s=s_1=\pm 2$
as a trivial,
we may use the auxiliary variable
$\Xi^{[K]}$ as defined
by the relation
 $s_1=\sqrt{ 2 \pm \Xi^{[K]}}$ at all the integers $K$.
The inspection of Table 4 then reveals that all the complete sets
of all the QE-compatiblity roots at all the listed integers $q=4K+3$
(i.e., at the set of $K = 1, 2, 3, 4$ and $ 5$)
coincide with the complete sets of roots of the
Chebyshev polynomials of the second kind,
 \be
 U_{K}
\left (
\frac{2-s_1^2}{2}
\right ) = 0, \ \ \ \ \
K = 1, 2, \ldots, K_0\,
\label{sed}
\ee
with the confirmed $K_0=5$ at present.
This means that in a way complementing the $N > N_0$
extrapolations performed in our paper PI we may now tentatively
extrapolate also the result (\ref{sed}) and conjecture that this
equation determines all the closed and exact solutions of our
$N=3$ QE Magyari-Schr\"{o}dinger eq. (\ref{trap}) also at {\em
all} the larger integers $q = 4K+3>4K_0+3=23$.

\subsection*{Acknowledgement}

Work supported by the grant Nr. A 1048302 of GA AS
CR.


\vspace{1cm}

\vspace{1cm}

\subsection*{The list of the table headings}

\begin{itemize}

\item
Table 1. Real QE roots $s=s_1$
 at the first few even~$q$ for $N=3$.

\item
Table 2.
Double Pascal triangle
for coefficients
 in the reduced secular equations
 $\sum_{k=0}^{q/2}\,s_1^k\,c_k^{(q)}=0$.
This extrapolates Table~1 to {\em all} the even $q$'s.

\item
Table 3.
 Real QE roots $s=s_1$
 at the first few odd~$q$ for $N=3$.

\item
Table 4.
 Real QE roots $s=s_1$
 at the first few  $q \equiv 3 \ (mod\ 4)$ for $N=3$.

\end{itemize}

\section*{Tables}

\vspace{1cm}

Table 1. Real QE roots $s=s_1$
 at the first few even~$q$ for $N=3$.

  $$
\begin{array}{||c||ccccc||} \hline \hline
q &  \multicolumn{5}{|c||}{{\rm
 roots\ }s=s_1 }\\
 \hline
 \hline
2&2&-1&&&\\
4&2&(\sqrt{5}-1)/2&-(\sqrt{5}+1)/2&&\\
6& 2& \multicolumn{4}{c||}{({\rm plus\ all\ three\ roots\ of\ }
a^3+a^2-2a-1)
}\\
8&2&-1&  \multicolumn{3}{c||}{({\rm plus\
all\ three\  roots\ of\ }a^3-3a+1)
}\\
10\ \ &2&  \multicolumn{4}{c||}{({\rm plus\ all\  five\ roots\ of\ }a^5
+a^4-4a^3-3a^2+3a+1)
}
\\
\hline
\hline
\end{array}
$$

 \newpage

Table 2.
Double Pascal triangle
for coefficients
 in the reduced secular equations
 $\sum_{k=0}^{q/2}\,s_1^k\,c_k^{(q)}=0$.
This extrapolates Table~1 to {\em all} the even $q$'s.

  $$
\begin{array}{||c||rrrrrrr||} \hline \hline
 \ \ &  \multicolumn{7}{|c||}{{\rm coefficients\ }c_k^{(q)} }\\
\hline
\hline
 \ \ \ \ k=&0&1&2&3&4&5&\ldots\\
 \ q \  &&&&&&&\\
 \hline
\ 2\ &1&1&&&&&\\
\ 4\ &-1&1&1&&&&\\
\ 6\ &-1&-2&1&1&&&\\
\ 8\ &1&-2&-3&1&1&&\\
\ 10\ &1&3&-3&-4&1&1&\\
\ 12\ &-1&3&6&-4&-5&1&\ddots \\
\ 14\ &-1&-4&6&10&-5&-6&\ddots\\
\ 16\ &1&-4&-10&10&15&-6&\ddots\\
\hline
\hline
\end{array}
$$

\vspace{1cm}


Table 3.
 Real QE roots $s=s_1$
 at the first few odd~$q$ for $N=3$.

  $$
\begin{array}{||c||cccccc||} \hline \hline
q &  \multicolumn{6}{|c||}{{\rm
roots\
}s=s_1 }\\
 \hline
 \hline
1&\ \ \ \  2
\ \ \ \  &-2&&&&\\
3&2&-2&&&&\\
5&2&1&-1& -2&&\\
7&2&\ \sqrt{2}\ &-\sqrt{2}\ & -2&&\\
9&2&\
 \sqrt{\frac{3 + \sqrt{5}}{2}}\ &\
 \sqrt{\frac{3 - \sqrt{5}}{2}}\ &\
- \sqrt{\frac{3 - \sqrt{5}}{2}}\ &\
- \sqrt{\frac{3 + \sqrt{5}}{2}}\ &-2\ \\
11&2&\ \sqrt{3}\ &1 & -1&-\sqrt{3}&-2\ \\
\hline
\hline
\end{array}
$$

\vspace{1cm}


Table 4.
 Real QE roots $s=s_1$
 at the first few  $q \equiv 3 \ (mod\ 4)$ for $N=3$.

  $$
\begin{array}{||c||ccc||} \hline \hline
q &  \multicolumn{3}{|c||}{{\rm
roots\
}s=s_1 }\\
 \hline
 \hline
3&\ \ \ \  \pm 2
\ \ \ \  &&\\
7&\pm 2&\pm \sqrt{2}&\\
11&\pm 2&\pm \sqrt{2 \pm 1}&\\
15&\pm 2&\pm \sqrt{2}&\pm \sqrt{2 \pm \sqrt{2}}\\
19&\pm 2&\pm \sqrt{2 \pm \left (
{\sqrt{5} \mp 1}
\right )/{2}} &\\
23&\pm 2&\pm \sqrt{2}&
\pm \sqrt{2 \pm \sqrt{2 \mp 1} } \\
\hline
\hline
\end{array}
$$

\end{document}